\documentclass[useAMS,usenatbib,usegraphicx]{mn2e}
\usepackage{times}

\title[]{Is Germanium (Ge, Z=32) A Neutron-Capture Element?}
\author[ Ping Niu, Weili Liu, Wenyuan Cui and Bo Zhang]{Ping Niu$^{1,2}$, Weili
Liu$^{1}$, Wenyuan Cui$^{1}$and Bo Zhang$^{1}$\thanks{Corresponding
author.
Email-address: zhangbo@mail.hebtu.edu.cn (Bo Zhang)}\\
$^{1}$Department of Physics, Hebei Normal University, 20 Nanerhuan
Dong Road, Shijiazhuang 050024, China\\
$^{2}$Department of Physics, Shijiazhuang  University, Shijiazhuang
050035, China}
\begin{document}

\date{Received...; in original form...}

\pagerange{\pageref{firstpage}--\pageref{lastpage}} \pubyear{2013}

\maketitle

\label{firstpage}

\begin{abstract}

Historically, Ge has been considered as a neutron-capture element.
In this case, the r-process abundance of Ge is derived through
subtracting the s-process abundance from total abundance of the
solar system. However, the Ge abundance of the metal-poor star HD
108317 is lower than that of the scaled ``residual r-process
abundance" in the solar system about 1.2 dex. In this work, based on
the comparison between the Ge abundances of the metal-poor star and
stellar yields, we find that the Ge abundances would not be produced
as the primary-like yields in the massive stars and mainly come from
the r-process. Based on the observed abundances of metal-poor stars,
we derived the Ge abundances of the weak r-process and main
r-process. The contributed percentages of neutron-capture process to
Ge in the solar system are about 59\%, which means that the
contributed percentage of Ge ``residual abundance" in solar system
is about 41\%. We find that the Ge``residual abundance" would be
produced as the secondary-like yields in the massive stars. This
implies that Ge element in the solar system is not only produced by
the neutron-capture process.

\end{abstract}

\begin{keywords}
nucleosynthesis, abundances - stars: abundances - stars: metal poor - element: germmanium
\end{keywords}

\section{Introduction}

Investigation of astrophysical origins of the elements is a
important task of modern astrophysics. The elements more heavier
than iron-group elements are mainly produced by slow neutron-capture
process (s-process) and rapid neutron-capture process (r-process)
\citep{Bu57}. The s-process contains two categories: the weak
s-process and the main s-process. The weak s-process mainly produces
the lighter neutron-capture elements (e.g., Sr) and occurs in
massive stars \citep{La97,Ra91,Ra93,Th00}. The contribution of
s-process to the heavier neutron-capture elements (e.g., Ba) is due
to the main s-process. The low-mass and intermediate-mass AGB stars
are usually considered to be the sites in which the main s-process
occur \citep{Bu99}. By subtracting the s-process abundances from the
solar system abundances, \citet{Ar99} derived the r-process
abundances, which are called as ``residual r-process abundances".

There are many evidences supporting that SNe II are the sites of the
r-process nucleosynthesis \citep{Co91,Sn08}. Because of the large
overabundance of r-process element Eu ([Eu/Fe]$\sim-1.6$), the two
``main r-process stars" CS 22892-052 and CS 31082-001 arouse special
attention: their abundance patterns of the heavier neutron-capture
elements fitted the  r-process pattern of solar system very closely
\citep{Co99,Tr02,Wa06,Sn08}. However, their lighter neutron-capture
elements (i.e., from Rb to Ag) are too defficient to agree with the
``residual r-process abundances" \citep{Sn00,Hi02}. This implies
that, for explaining the r-process abundances of the solar system,
another process, referred to as the ``lighter element primary
process" \citep{Tr04} or ``weak r-process" \citep{Is05}, is
required.

By comparison, the abundance patterns of weak r-process stars, HD
122563 and HD 88609, show excess of lighter neutron-capture
elements and defficiency of heavier neutron-capture elements,
which are obviously different from the patterns of the main
r-process stars \citep{We00,Jo02,Ho07}. Based on the abundance
analyze of metal-poor stars, \citet{Mo07} found that the weak
r-process abundance pattern is uniform and unique. \citet{Kr07}
have performed the r-process calculations. They found that the
main r-process abundances can only be matched under the conditions
of high neutron densities ($23<logn_{n}<28$). On the other hand,
their calculations indicate that smaller neutron number densities
($20<log n_{n}<22$) could reproduce the weak r-process abundances.
Recently, \citet{Li13a} found that the abundances of
neutron-capture elements in all metal-poor stars, including main
r-process stars and weak r-process stars, contain the
contributions of two r-processes.

The element germanium is in the transition between the iron group
elements and neutron-capture elements. To date, the quantitative
contributions of germanium (Ge, $Z=32$) from various astrophysical
scenarios are rarely known in the studied metal-poor stars.
Traditionally, Ge ($Z = 32$) has been considered as one of the
lightest neutron-capture elements \citep{Sn98}. It is thought to be
produced through the s-process and r-process. Based on chemical
evolution calculations, \citet{Tr04} reported that about 12\% of Ge
abundance in the solar system is produced in the AGB stars. On the
other hand, about 43\% of the Ge abundance has been produced by the
weak s-process via neutron captures through the
$^{22}Ne(\alpha,n)^{25}Mg$ reaction in massive stars \citep{Ra92}
and around 45\% of the Ge abundance belong to `` residual r-process
abundance" \citep{Tr04,Sn08}. Note that the r-process abundances in
solar system are derived through subtracting the s-process
abundances from total abundances of solar system \citep{Ar99,Si04}.

Adopting updated neutron-capture cross sections, \citet{Pi10}
presented new weak s-process calculations for a massive star with 25
M$_{\odot}$ at solar metallicity. They found that Ge is one of the
most abundant s-elements in the He core and the C shell of the
massive star and speculated that the weak s-process is responsible
for about 80\% of Ge abundance in the solar system. However, they
point out that their calculated results cannot be regarded as the
weak s-component of the solar system, because the results come from
only one stellar model. They suggested that, for deriving the weak
s-component, the contributions from massive stars with various
masses must be considered. In this case, the averaged yields
weighted by the initial mass function (IMF) should be more effective
than the yields of individual massive star (e.g.,\citet{Ra93}).
Furthermore, based on the observed [Ge/Fe] ratios of the metal-poor
stars, they estimated that the contribution from the primary-like
yields produced in massive stars to the solar germanium is about
5\%-8\%. Recently, \citet{Fr12} reported that large variations in
weak s-process yields could occur since the rotating of the
metal-poor massive stars.

Based on the abundance analysis of neutron-capture elements for
metal-poor stars, \citet{Co05} found that the abundances of the
third r-process peak elements correlated with the abundances of the
r-process element Eu, implying a common origin or site for r-process
nucleosynthesis of heavier (Z $>$ 56) elements. On the other hand,
the Ge abundances correlated with the Fe abundances well. They
estimated that the observed relation is [Ge/H]$=$ [Fe/H]$-0.79$,
which is lower than the ratio of the solar system about 0.8 dex. The
abundance relation between Ge and Fe would mean that an explosive
process or charged-particle process, rather than neutron-capture
process, would be responsible for the production of Ge at low
metallicities \citep{Co05}. Recently, \citet{Ro12a} and
\citet{Ro12b} analyzed the abundances of the metal-poor star HD
108317 and found that the abundances can not be matched by the
s-process pattern or by the scaled ``residual r-process" pattern of
the solar system. HD 108317 shows overabundance of the heavier
neutron-capture elements from Eu to Pt, but the discovery that Ge,
which is one of the lightest neutron-capture element, is deficient
is puzzling. The Ge abundance of HD 108317 is about 1.2 dex lower
than that of scaled  Ge ``residual r-process abundance" in the solar
system. They proposed that the elements heavier than Ge (i.e., As
and Se) should be the point where the r-process turns on.
\citet{Si13} point out that the observed Ge abundance in metal-poor
stars should serve as the key conditions in constraining the
r-process nucleosynthesis. Obviously, understanding of the
astrophysical origins of Ge abundance in our Galaxy is a challenging
task \citep{Ro12b}.

In this paper, based on the abundance approach for metal-poor
stars presented by \citet{Li13a}, we calculate the relative
contributions from the individual neutron-capture process to the
abundances of metal-poor stars in which Ge abundances have been
observed. In Section 2, we extract abundance clues of the weak
r-process and main r-process for Ge from weak r-process star HD
122563 and main r-process star CS 31082-001.  In Section 3, based
on the observed abundances, we derived the abundances of weak
r-process and main r-process for Ge element. The calculated
results are discussed in Section 4 and Section 5. In Section 6,
the astrophysical origin of ``residual abundance" of Ge was
investigated. Our conclusions are given in Section 7.

\section{ABUNDANCE CLUES}

The element germanium is in the transition between charged-particle
synthesis of the iron-group elements and neutron-capture synthesis
of heavier elements. There are 19 stars that have been observed for
Ge abundance. The values of log$\varepsilon$(Ge),
log$\varepsilon$(Fe) and log$\varepsilon$(Eu)
\citep{Co05,Ro12a,Ro12d,Si13,We00}
for the sample stars are listed in Table 1. The standard definitions
of elemental abundances and abundance ratios are used throughout the
paper. For element X, the abundance is defined as the logarithm of
the number of atoms of element X per $10^{_{12}}$ hydrogen atoms,
log$\varepsilon(X)$=log$(N_{X}/N_{H})$+12.0. The
abundance ratio relative to the solar ratio of element X and element
Y is defined as [X/Y]=log($N_{X}/N_{Y}$)-log($N_{X}/N_{Y})_{\odot}$.
The ratios of [Ge/H], [Fe/H] and [Eu/H] have been rescaled to
corresponding solar logarithm abundances of
log$\varepsilon$(Fe)=7.50, log$\varepsilon$(Eu)=0.54 and
log$\varepsilon$(Ge)=3.62 \citep{An89}. In this section, we wish to
compare the observed Ge abundances of the metal-poor stars which are
formed in the interstellar medium (ISM) with the yields of the
massive stars to investigate the astrophysical origin of Ge. In this
case, the averaged yields weighted by IMF should be more effective
than the yields of individual massive star. The solid lines in
Figure 1 (a) and (b) show the averaged yields of massive star with a
mass range from 10 M$_{\odot}$ to 100 M$_{\odot}$ for zero
metallicity (standard mixing of 0.1, explosion energy of 5.0
($\times10^{51}$ergs)) presented by \citet{He10}, which have been
weighted by the Salpeter initial mass function (IMF) and scaled to
the Fe abundances of the weak r-process star HD 122563 and main
r-process star CS 31082-001, respectively. The observed elemental
abundances of HD 122563 \citep{Ho07,We00} and CS 31082-001
\citep{Hi02,Si13} marked by filled circles are also shown to
facilitate comparison. The scaled IMF-weighted yields of massive
stars with a mass range from $10 M_{\odot}$ to $50 M_{\odot}$ for
zero metallicity presented by \citet{Ko06} (see their Table 3) are
also plotted in Figure 1(a) and (b) by dashed lines. From the figure
we can see that the scaled Ge abundances are lower than the Ge
abundances of the weak r-process star HD 122563 ([Ge/Fe]=-0.9) and
main r-process star CS 31082-001 ([Ge/Fe]=-0.63) about 1.0 dex and
1.3 dex respectively. Note that the ratio of [Ge/Fe]=-0.9 listed in
Table 1 corresponds to that the scaled Ge abundance is lower than
the observed Ge abundance about 1.0 dex. Because the observed ratios
of [Ge/Fe] in the metal-poor stars ([Fe/H]$\leq-1.8$) lies in the
range of $-0.63\sim-1.2$ \citep{Co05}, for the other metal-poor
stars, the scaled Ge abundances produced in the massive stars are
lower than the observed Ge abundances about 0.7 dex at least. The
compared results mean that the Ge abundances in the metal-poor stars
would not be produced as the primary-like yields in the massive
stars. Obviously, another process, which is responsible for the Ge
abundances in the metal-poor stars, is needed. Because HD 122563 is
a weak r-process star and CS 31082-001 is a main r-process star,
their Ge abundances, similar to the abundances of other
neutron-capture elements, should mainly come from the r-process.

\begin{table*}
 \centering
\begin{minipage}{100mm}
\caption{The values of log$\varepsilon$(Ge I), log$\varepsilon$(Fe
I), log$\varepsilon$(Fe II) and log$\varepsilon$(Eu II) }
  \begin{tabular}{@{}lcccccccccr@{}}
  \hline
 Star & log$\varepsilon$(Ge I)   & log$\varepsilon$(Fe I)
&log$\varepsilon$(Fe II)&log$\varepsilon$(Eu II) & References \\

\hline

sun     &   3.63    &   7.5 &   7.5 &   0.54    &   1   \\
HD 108317   &   0.17    &   4.97    &   5.13    &   -1.37   &   2   \\
HD 126587   &   0.03    &   4.57    &   4.69    &   -1.89   &   3   \\
HD 186478   &   0.35    &   4.94    &   5.06    &   -1.5    &   3   \\
HD 221170   &   0.69    &   5.15    &   5.47    &   -0.85   &   3   \\
BD +17$^{\circ}$3248    &   0.46    &   5.42    &   5.4 &   -0.67   &   3   \\
HD 6268     &   0.32    &   5.08    &   5.14    &   -1.33   &   3   \\
HD 128279   &   -0.03   &   5.02    &   5.04    &   -1.96   &   2   \\
HD 126238   &   0.62    &   5.52    &   5.57    &   -1.19   &   2   \\
HD 122956   &   0.84    &   5.55    &   5.81    &   -0.79   &   3   \\
HD 6755     &   1.08    &   5.82    &   5.93    &   -0.5    &   3   \\
HD 115444   &   -0.05   &   4.54    &   4.52    &   -1.63   &   4   \\
    &   -0.07   &   4.6 &   4.79    &   -1.61   &   3   \\
CS 31082-001    &   0.1 &   4.6 &   4.58    &   -0.75   &   5   \\
HD 122563   &   -0.03   &   4.78    &   4.78    &   -2.59   &   4   \\
    &   -0.16   &   4.78    &   4.89    &   -2.59   &   3   \\
HD 175305   &   1.28    &   6.02    &   6.14    &   -0.29   &   3   \\
HD 94028    &   1.56    &   5.75    &   5.88    &   -0.88   &   6   \\
HD 76932    &   2.4 &   6.32    &   6.58    &   -0.13   &   6   \\
HD 107113   &   2.54    &   6.71    &   6.97    &   $ <$0.4 &   6   \\
HD 2454     &   2.7 &   6.78    &   7.06    &   0.04    &   6   \\
HD 16220    &   3.11    &   6.94    &   7.09    &   0.14    &   6   \\
HD 140283   &   $ <$0.6      &   4.79    &   4.88    &   $ <$-1.7 &   6     \\
CS 22892-052&   $ <$-0.2     &   4.4     &   4.41    &   -0.95 &   3    \\
HD160617    &   $ <$1.2    &   5.58    &   5.73    &   -0.81 &   7      \\

\hline

\end{tabular}

References.-1 Anders \& Grevesse1989; 2 Roederer et al.2012a; 3 Cowan et al. 2005;
4 Westin et al.2000; 5 Siqueira Mello et al.2012;6 Roederer 2012d; 7 Roederer et al.2012b

\end{minipage}
\end{table*}

\begin{figure*}
 \centering
 \includegraphics[width=1.0\textwidth,height=0.30\textheight]{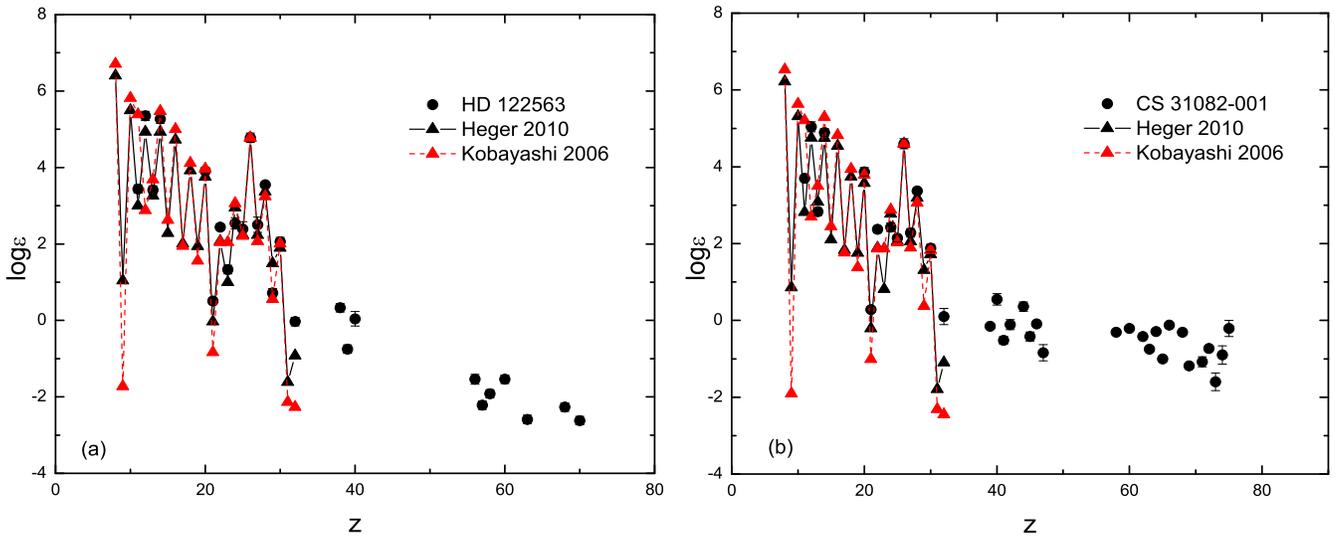}
\caption{The solid lines show the averaged yields of the massive
stars with zero metallicity presented by \citet{He10}, which have
been weighted by the Salpeter IMF and scaled to the Fe abundances of
the weak r-process star HD 122563 (a) \citep{We00} and main
r-process star CS 31082-001 (b) \citep{Hi02,Si13}, respectively. The
deshed lines in Figure1(a) and (b) show the IMF-weighted yields of
massive star with zero metallicity presented by \citet{Ko06} and
also scaled to the Fe abundances of the two stars. The observed
elemental abundances marked by filled circles are also shown to
facilitate comparison.}

\end{figure*}
\section{WEAK R-PROCESS AND MAIN R-PROCESS ABUNDANCES OF GE}

\citet{Tr99} have found that the main s-process contributions are
significant starting from [Fe/H]$\simeq-1.5$. On the other hand, for
the weak s-process, little contribution is expected in halo stars
\citep{Tr04}, because of the strong decrease in its efficiency with
decreasing metallicity. These results imply that the abundances of
neutron-capture elements in the metal-poor stars dominantly come
from the r-process. For exploring the origin of the Ge abundances in
metal-poor stars quantitatively, we adopt the abundance approach
presented by \citet{Li13a}. The abundance of the ith element can be
calculated as:

\begin{equation}
N_{i}(Z)=(C_{r,m}N_{i,r,m}+C_{r,w}N_{i,r,w})\times{10^{[Fe/H]}}
\end{equation}
where $N_{i,r,m}$ is the abundance produced by the main r-process,
and $N_{i,r,w}$ is the abundance produced by the weak r-process,
whereas $C_{r,m}$ and $C_{r,w}$ are the corresponding component
coefficients.

For weak r-process star HD 122563, the r-process component
coefficients are $C_{r,m} =0.26$ and $C_{r,w}= 4.07$. The
corresponding values of main r-process star CS 31082-001 are
$C_{r,m}=52.04$ and $C_{r,w} =3.89$ \citep{Li13a}. Based on the
observed Ge abundances of HD 122563 \citep{We00} and CS 31082-001
\citep{Si13}, from equation (1), the values of $N_{Ge,r,w}$and
$N_{Ge,r,m}$ can be derived as:

\begin{equation}
N_{Ge,r,m}=0.38;
N_{Ge,r,w}=3.87.
\end{equation}

Li et al.(2013) have used the ``percentage of weak r-process"
$f^{r}_{r,w}$ (i.e., $N_{i,r,w}/(N_{i,r,m} + N_{i,r,w}))$ and the
``percentage of main r-process" $f^{r}_{r,m}$(i.e.,
$N_{i,r,m}/(N_{i,r,m} +N_{i,r,w}))$ to calculate the relative
contributions to r-process abundances in the solar system. They
found that there are the linearity decrease trend in $f^{r}_{r,w}$
from atomic number Z=30 to Z=63. Based on our calculation,
``percentage of weak r-process" $f^{r}_{r,w}$=0.91 for Ge, which is
higher than ``percentage of weak r-process" for Sr, Y and Zr
($\sim$0.7). Our calculated results are consistent with the decrease
trend in the ``percentage of weak r-process" $f^{r}_{r,w}$. Further
more, our result is also consistent with \citet{Is05}, who found
that the contribution from weak r-process to the r-process
abundances decreases with increasing atomic number.

\begin{figure*}
 \centering
 \includegraphics[width=0.60\textwidth,height=0.35\textheight]{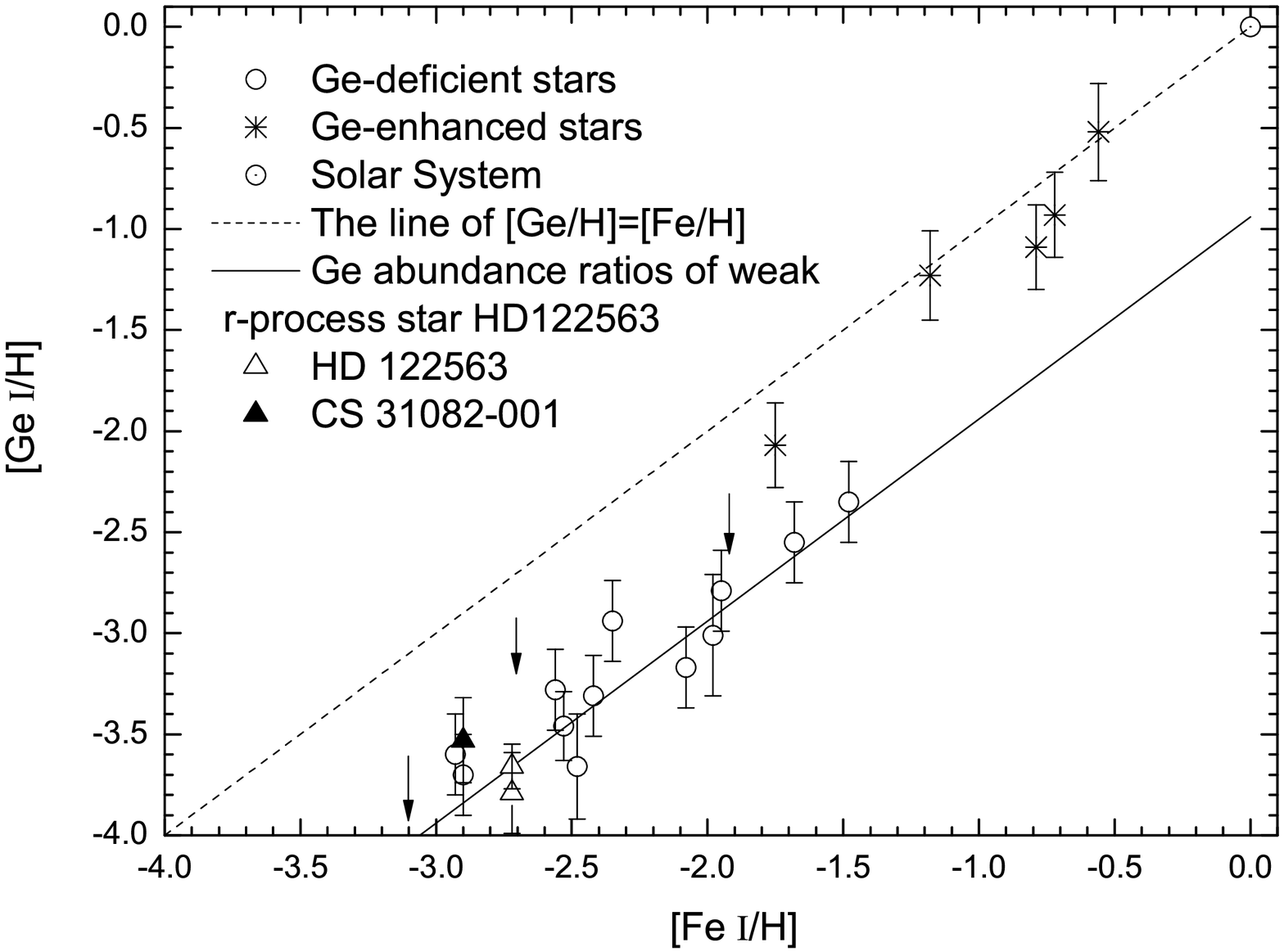}
\caption{The value of the logarithmic ratio [Ge/H] displayed as a function
of [Fe/H] metallicity. The dashed line and solid line indicate the
Ge abundance ratios of solar system and weak r-process star HD122563,
respectively. Open circle are the abundance ratios of Ge-deficient
stars adopted from \citet{Co05,Si13}. Asterisk are abundance ratios
of Ge-enhanced stars from \citet{Ro12d}. The arrows represent
the upper limits for CS 22892-052 ([Fe/H]=-3.1), HD 160617 ([Fe/H]=-1.92) and HD 140283 ([Fe/H]=-2.71).}
\end{figure*}

\begin{table*}
 \centering
 \begin{minipage}{100mm}
\caption{The abundances and contributed  percentages of main r-,
weak r-, main s-, weak s- and ``residual process"  to Ge in the
solar system.}

  \begin{tabular}{@{}lccccclrc@{}}
  \hline
 Element & Z &$N_{i,total}$
&$N_{i,r,m}$ &$N_{i,r,w}$  &$N_{i,s,m} $  &$N_{i,s,w}$ &$N_{i,res}$\\

 \hline
Ge         &  32  & $119^{a}$  & 0.38  & 3.87 & $ 14.3^{b}$  & $51.20^{c}$ & 49.30\\
percentage ( \% ) && 100  & 0.32  & 3.25 &  12.0  & 43.0  & 41.43\\
\hline
\end{tabular}

\end{minipage}

$^{a}$ Anders \& Grevesse 1989
$^{b}$ Travaglio et al.2004
$^{c}$ Raiteri et al. 1992
\end{table*}

In the early Galaxy, a part of iron group elements and light
elements were produced in first generations of very massive stars.
The corresponding abundance pattern was called as the prompt (P)
component  \citep{Qi01}. \citet{Li13a} derived the P-component from
the low-[Sr/Fe] star BD-18 5550. Considering the contribution from
P-component, the r-process component and P-component coefficients of
weak r-process star HD 122563 are $ C_{r,m} = 0.26$, $ C_{r,w} =
3.63$ and $C_{p}=0.16$, respectively. The r-process component and
P-component coefficients of main r-process star CS 31082-001 are $
C_{r,m} = 3.23$, $ C_{r,w} = 52.32 $ and $C_{p}=0.15 $ and the
corresponding values of metal-poor star HD 115444 are $C_{r,m} =
7.98$, $C_{r,w} = 4.90 $ and $C_{p}=0.0 $ \citep{Li13a},
respectively. Based on the observed Ge abundances of HD 122563
\citep{We00}, CS 31082-001 \citep{Si13} and HD 115444 \citep{Co05},
from equation (4) presented by \citet{Li13a}, the values of
$N_{Ge,r,w}$, $N_{Ge,r,m}$ and $N_{Ge,P}$ can be derived as:
$N_{Ge,r,m}= 0.296$; $N_{Ge,r,w} = 4.29$; $N_{Ge,P}<0.0$. As a test,
using Ge abundance of another metal-poor star HD 221170 \citep{Co05}
and its component coefficients \citep{Li13a} to replace those of HD
115444, we also derived $N_{Ge,P}<0.0$. Our calculated results imply
that the contributions from P-component to Ge abundance for the
metal-poor stars could be negligible. \citet{Li13a} have found that
the contributions of the P-component to the light elements and iron
group elements are observable only for the low-[Sr/Fe] star
([Sr/Fe]$<-1.0$). Because there is no low-[Sr/Fe] star in the sample
stars, the contributions of P-component are not included in our
calculation.

The main r-process abundance $N_{Ge,r,m}$ and weak r-process
abundance $N_{Ge,r,w}$ are listed in Table 2. For comparison, the
solar abundance $N_{Ge,total}$ \citep{An89}, main s-process
abundance $N_{Ge,s,m}$ \citep{Tr04} and weak s-process abundance
$N_{Ge,s,w}$ \citep{Ra92} are listed in Table 2, respectively. It is
important to note that the solar Ge abundance is higher than the sum
of $N_{Ge,r,w}$, $N_{Ge,r,m}$, $N_{Ge,s,m}$ and $N_{Ge,s,w}$.
Namely, there is ``residual abundance" of Ge, which is listed in
Table 2, between the abundance of solar system and the sum of
contributions from all neutron-capture processes. The contributed
percentages of main r-, weak r-, main s-, weak s- process and
``residual abundance" of Ge to solar system are 0.32\%, 3.25\%,
12.0\%, 43.0\% and 41.43\% respectively, which are also listed in
Table 2. Note that the contributed percentage of Ge ``residual
abundance"  to solar system is larger than 40\% and the Ge
``residual abundance" do not come from neutron-capture processes.
\citet{An89} reported that the nuclear statistical equilibrium (NSE)
process should be responsible for the production of Ge partly. The
astrophysical site related to the Ge residual abundance will be
investigated in Section 6. There is a fact that the Ge abundances of
the metal-poor stars are lower than the scaled ``residual r-process
abundance" in the solar system \citep{Ro12a} about 1.0 dex. From
Table 2 we can see that the contributed percentage of the sum of the
r-process for Ge is only about 4\%, which is smaller than that of
the ``residual r-process abundance" in the solar system about one
order of magnitude. This should be the astrophysical reason that the
Ge abundances of metal-poor stars are lower than that of the scaled
``residual r-process abundance" in the solar system.

In order to investigate the origin of Ge, [Ge/H] as a function of
metallicity is particularly useful \citep{Co05,Si13}. The variation
of the logarithmic ratio [Ge/H] with metallicity is shown in Figure
2. The dashed line and solid line indicate the Ge abundance ratios
of solar system (i.e., [Ge/H]=[Fe/H]) and weak r-process star
HD122563, respectively. It is clearly that the ratio [Ge/H] of the
weak r-process star is lower than that of the solar system about 0.9
dex. We can see that for the most metal-poor stars, the observed
[Ge/H]$\sim$[Fe/H] correlation is close to the line of abundance
ratio of weak r-process star HD122563, which is close to the
findings of \citet{Co05}. The results imply that the Ge abundances
in these stars mainly come from the weak r-process. Note that, for
the main r-process star CS 31082-001, the ratio [Ge/H] is higher
than the line of abundance ratio of weak r-process star about 0.3
dex, which should be the evidence that its Ge abundance partly come
from main r-process material. Based on the ratios [Ge/H], the sample
stars revealed the existence of two distinct groups, i.e.,
Ge-enhanced stars having [Ge/H]$\geq-2.4$ and Ge-deficient stars
having [Ge/H]$\leq-2.5$. The Ge-deficient stars should be formed in
regions mainly polluted by r-process material. From the figure we
can see that, for the Ge-enhanced stars, the ratios [Ge/H] are close
to the solar abundance ratio. Their Ge abundances would reflect the
contributions of various processes, including the r-process,
s-process and an unknown astrophysical process. The unknown process
would be responsible for the ``residual abundance" of Ge in the
solar system.

For providing a more direct comparison of Ge abundance to the main
r-process abundance, Figure 3 shows the abundance ratios [Ge/H] as a
function of the abundance ratios [Eu/H]. The ratios [Ge/H] for the
most Ge-deficient stars are lower than the solar ratio about 1.4 dex
at similar [Eu/H]. We can see that the Ge abundance ratios of most
metal-poor stars, except for the weak r-process star HD 122563, are
lower than that of the scaled ``residual r-process abundance" of Ge
in the solar system at least 0.8 dex, which suggests that the Ge
deficient is a common phenomenon in the metal-poor stars, agreeing
with the finding of \citet{Ro12a}. For the most Ge-deficient stars,
the ratios [Ge/H] increase with increasing [Eu/H], except for the
weak r-process star HD 122563 and main r-process star CS 31082-001.
It is very interesting to note that, for the most Ge-deficient
stars, the ratios [Ge/H] are close to the r-process abundance ratio
in the solar system, which is plotted by solid line in Figure 3. The
result is consistent with the suggestion that the Ge abundances in
the metal-poor stars mainly come from the r-process. Furthermore,
from the figure we can see that the ratio [Ge/H] of the main
r-process star CS 31082-001 is lower than the ratio of solar about
2.0 dex, since the main r-process produce large amount of main
r-process element Eu and small amount of Ge. On the other hand,
because the weak r-process produce larger amount of Ge and hardly
produce Eu, the ratio [Ge/H] of the weak r-process star HD 122563 is
lower than the ratio of solar only about 0.3 dex and close to the
ratio of ``residual r-process abundance" in the solar system.
Obviously, the abundance ratios of Ge-enhanced stars are higher than
the r-process abundance ratio in the solar system, which should be
the evidence that their Ge abundances have contained the
contributions of the s-process and other astrophysical process.

\section{ CALCULATED RESULTS ABOUT THE GE-DEFICIENT STARS}

Based on Equation (1), using the observed data of the Ge-deficient
stars \citep{Ho04,Mi01,Ro10,Iv06,Co02,Co05, Ro12a}, we can obtain
the best-fit $C_{r,m }$ and $C_{r,w}$ by looking for the minimum
$\chi^{2}$. The component coefficients and $\chi^{2}$ are listed in
Table 3. As an example, Figure 4(a) shows the best-fit results for
the sample star HD 221170. The observed elemental abundances marked
by filled circles are also shown to facilitate comparison.

The Ge-deficient star HD 108317 shows enrichment by the s-process
\citep{Ro12c}. Considering the s-process contribution, the ith
element abundance can be calculated as:

\begin{equation}
N_{i}(Z)=(C_{r,m}N_{i,r,m}+C_{r,w}N_{i,r,w}+C_{s,m}N_{i,s,m})\times{10^{[Fe/H]}}
\end{equation}
where $N_{i,s,m}$ is the abundance produced by the main s-process in
the AGB star and $C_{s,m}$ is the main s-process component
coefficient (see Sect 5 for details on the s-processes). The adopted
abundance $N_{i,s,m}$ in Equation (3) is taken from the main
s-process abundance with [Fe/H]$=-2.6$ for $1.5M_{\odot}$ given by
\citet{Bi10} (see their Table B6) , which has been normalized to the
s-process abundance of Ba in the solar system \citep{Ar99}. The
component coefficients are listed in Table 3 and fitted results are
presented in Figure 4(b). In the top panel of Figure 5, the
individual offsets $\bigtriangleup\log\varepsilon$ for
12 Ge-deficient stars compare to the calculated results are shown.
The r.m.s. offsets in $\log\varepsilon$ are shown in the bottom
panel. Typical observational errors are $0.2\sim 0.3$ dex (dotted
lines). It could be found from Figure 5 that the individual relative
offsets for most elements are smaller than 0.30 dex and the
root-mean-square offsets are close to zero. It is clear from Figure
4 and Figure 5 to confirm the validity of the values of $N_{Ge,r,w}$
and $N_{Ge,r,m}$ derived in this work.

\begin{figure*}
 \centering
 \includegraphics[width=0.6\textwidth,height=0.35\textheight]{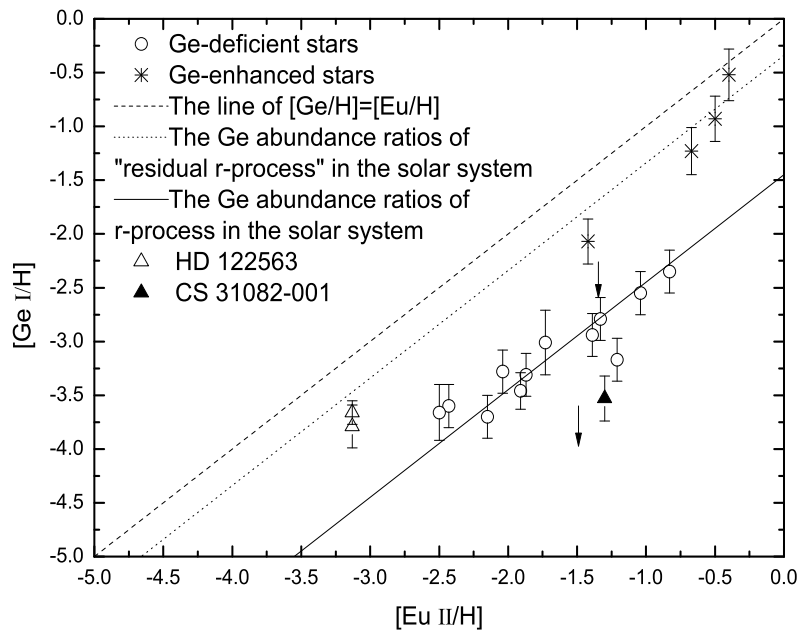}
\caption{The [Ge/H] ratios as a function of [Eu/H]. The dashed line shows
the solar abundance ratio of Ge, the dotted line indicates the solar
``residual r-process abundance" ratios of Ge, while the solid line indicates
the solar r-process abundance ratios of Ge. The observed data as in Figure 2.
The arrows represent the upper limits for CS 22892-052 ([Eu/H]=-1.49)
and HD 160617 ([Eu/H]=-1.35).}

\end{figure*}

\begin{table*}
 \centering
 \begin{minipage}{85mm}
  \caption{The component coeffcients of the two r-processes, main s-process
 and $\chi^{2}$ and $K- K_{free}$ for Ge-deficient
stars. ($\log{\varepsilon}=\log{N}+1.55$)\label{tbl-2}}

  \begin{tabular}{@{}lcccccrrc@{}}
  \hline
 Star & [Fe/H] &$C_{r,m}$
 & $C_{r,w}$ & $C_{s,m}$ & $\chi^{2}$   &  $K- K_{free}$\\

 \hline

HD 108317   &   -2.53   &   4.3 &   3.5 &   0.016   &   0.84    &   30  \\
HD 126587   &   -2.93   &   5.2 &   5.9 &       &   1.47    &   26  \\
HD 186478   &   -2.56   &   4.8 &   4.7 &       &   0.71    &   28  \\
HD 221170   &   -2.35   &   9.1 &   5.5 &       &   1.11    &   41  \\
BD +17$^{\circ}$3248    &   -2.08   &   10.2    &   3.8 &       &   2.59    &   35  \\
HD 6268     &   -2.42   &   3.2 &   2.4 &       &   1.92    &   29  \\
HD 128279   &   -2.48   &   0.8 &   2.9 &       &   2.06    &   22  \\
HD 126238   &   -1.98   &   2.2 &   3.3 &       &   0.4 &   30  \\
HD 122956   &   -1.95   &   7.2 &   5.1 &       &   0.52    &   9   \\
HD 6755     &   -1.68   &   6.1 &   4.2 &       &   0.3 &   9   \\
HD 115444   &   -2.9    &   6   &   4.1 &       &   1.06    &   29  \\
HD 175305   &   -1.48   &   2.1 &   2.2 &       &   1.26    &   31  \\

\hline
\end{tabular}
\end{minipage}
\end{table*}

\citet{Si13} found a clear anticorrelation between the Ge
enhancement and the r-process richness. Figure 6 shows the
homogenized  Ge abundances relative to the level of the main
r-process element Eu as a function of the enrichment in r-process
element Eu. The dash line corresponds to the abundance ratio of weak
r-process star and dotted line represents the abundance ratio of the
main r-process component. We find that the observed abundance ratios
are close to the dash line for the sample stars with [Eu/Fe]$\leq$
1.0. However, the observed abundance ratios are close to the main
r-process line but not weak r-process line for the sample star(s)
with [Eu/Fe]$>$1.0. At low metallicities the abundances can be well
described by a mixture of two r-processes. We can adopt the
abundances of weak r-process star HD 122563 as the initial
abundances of Ge-deficient stars, since its abundances contain the
lowest contribution of the main r-process. The abundance of ith
element in the Ge-deficient stars can be calculated as
\begin{equation}
N_{i}=N_{i,HD 122563}+C_{r,m}N_{i,r,m}
\end{equation}
where $N_{i,HD 122563}$ is the abundance of weak r-process star HD
122563. To find an explanation of the relation between [Ge/Eu] and
[Eu/Fe] quantitatively, taking $C_{r,m}$ ranging from 0.2 to 200,
based on equation(4), in Figure 6, the solid line correspond to the
abundance ratios enriched by main r-process material. For the
metal-poor stars with [Eu/Fe]$\leqslant$1.0, the Ge abundances
dominantly come from weak r-process. The increasing of [Eu/Fe]
directly leads to a decline of [Ge/Eu]. Furthermore, based on the
calculation, [Ge/Eu] flattened for higher [Eu/Fe], since the
contributions from the main r-process component to Ge increase. We
can see that, for the Ge-deficient stars, the observed
[Ge/Eu]$\sim$[Eu/Fe] relation can be explained by the calculated
results. From the figure we can see that the Ge abundances of the
Ge-enhanced stars fall above the r-process line, whose astrophysical
reasons will be studied in Section 5 and Section 6.

Based on the abundance analysis of neutron-capture element, we found
that the Ge abundances of Ge-deficient stars mainly come from the
weak r-process, except for the main r-process stars. Although light
and iron group elements are not produced by weak r-process
nucleosynthesis, both iron group elements and the weak r-process
elements at low metallicity would be produced in the massive stars
\citep{Li13a}. The Ge yield and Fe yield possess primary nature
(i.e., yields that are not effect by the initial metallicity
approximately), which contributed Ge and iron group elements to ISM
for various metallicity. So the abundances of iron group elements
could be correlated closely with those of the weak r-process
elements in Ge-deficient stars.  In this case, the observed
correlation of the [Ge/H] with the [Fe/H] shown in Figure 2, which
had been found by \citet{Co05}, could be understood.

\begin{figure*}
 \centering
 \includegraphics[width=1.0\textwidth,height=0.3\textheight]{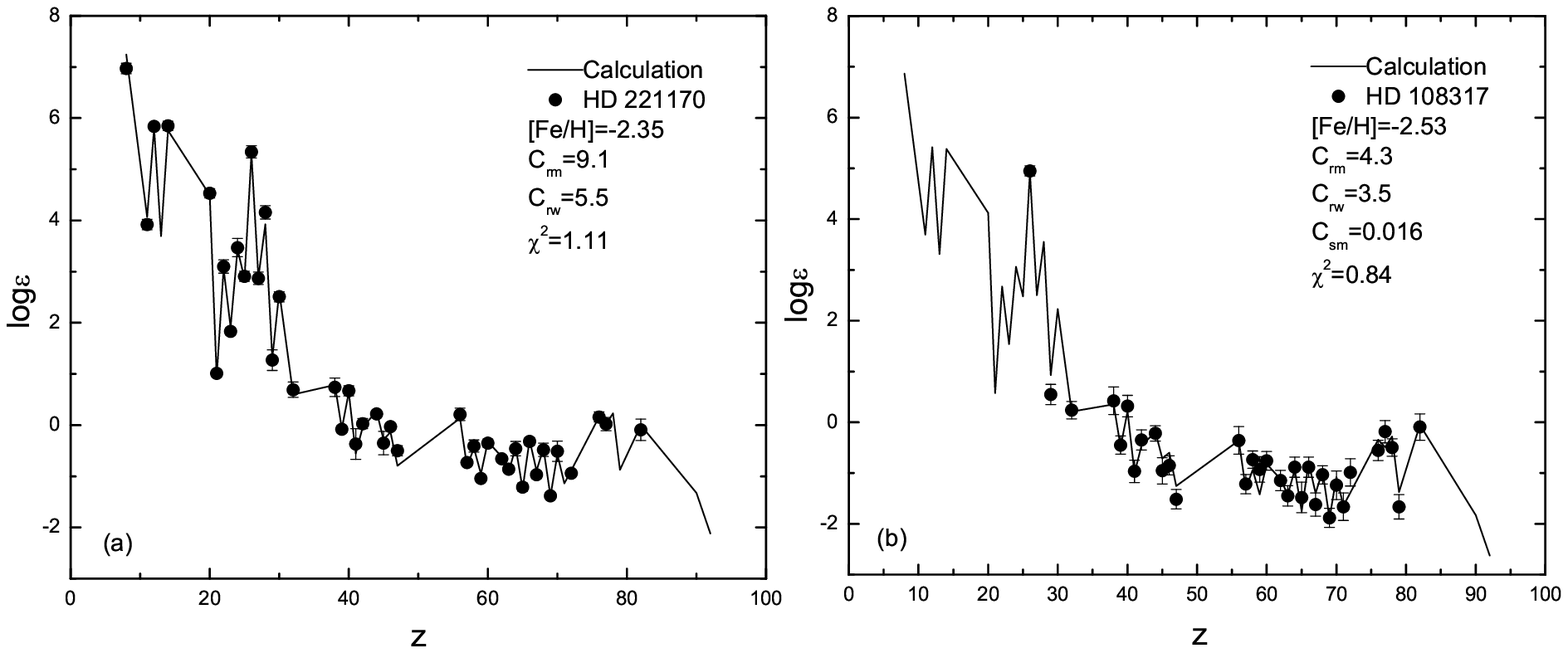}
\caption{Best fitted results of 2 sample stars. The filled circles
with error bars are the observed element abundances, the solid line
represents the calculated results.}

\end{figure*}

\begin{figure*}
 \centering
 \includegraphics[width=0.58\textwidth,height=0.35\textheight]{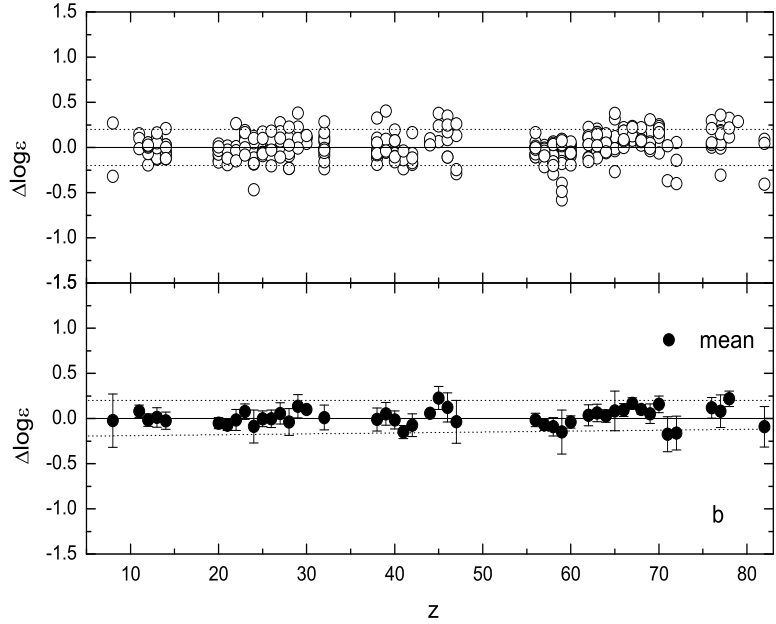}
\caption{Top panel: the individual relative offsets
($\Delta\log\varepsilon(X)\equiv\Delta\log\varepsilon(X)_{cal}-\Delta\log\varepsilon(X)_{obs}$)
for the Ge-deficient stars with respect to the predictions from the
abundance approach. Bottom panel: the root-mean-square offset of
these elements in $log\varepsilon$, typical observational
uncertainties in $log\varepsilon$ are $\sim0.2-0.3$ dex (dotted
lines).}

\end{figure*}

\begin{figure*}
 \centering
 \includegraphics[width=0.6\textwidth,height=0.35\textheight]{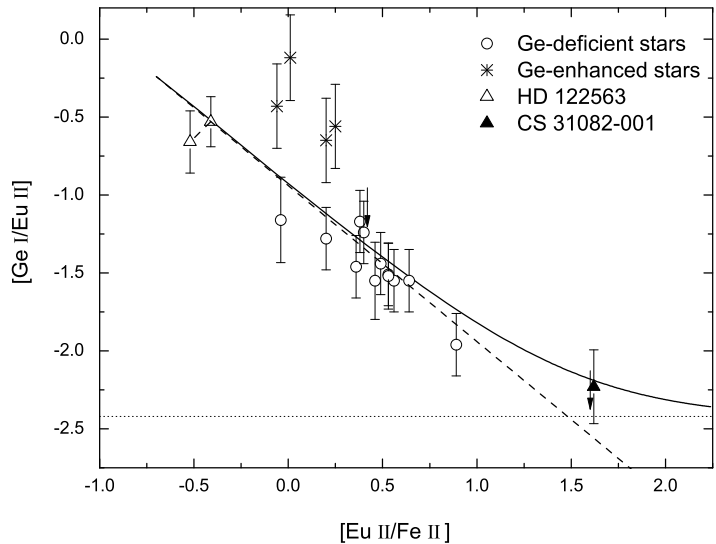}
\caption{The observed abundances of Ge relative to Eu as a function
of the abundance ratios of Eu are shown. The solid line corresponds
to the abundance ratios enriched by a mixture of both weak and main
r-process material. The dash line corresponds to the abundance ratio
of weak r-process star and dotted line represents the abundance
ratio of the main r-process component. The observed data as in
Figure 2. The arrows represent the upper limits for CS 22892-052
([Eu/Fe]=1.6) and HD 160617 ([Eu/Fe]=0.42).}

\end{figure*}

\begin{table*}
 \centering
 \begin{minipage}{82mm}
  \caption{The component coeffcients of the two r-processes, two s-process,
  and $\chi^{2}$ for sample stars. ($\log{\varepsilon}=\log{N}+1.55$)\label{tbl-2}}

  \begin{tabular}{@{}lccccclrc@{}}
  \hline
 Star & [Fe/H] &$C_{r,m}$ & $C_{r,w}$ & $C_{s,m}$ & $C_{s,w}$ & $\chi^{2}$  \\

 \hline

HD 108317   &   -2.53   &   4.3 &   3.5 &   0.016   &   0   &   0.86    \\
HD 126587   &   -2.93   &   5.2 &   5.9 &   0.02    &   0   &   1.58    \\
HD 186478   &   -2.56   &   4.8 &   4.7 &   0   &   0   &   0.76    \\
HD 221170   &   -2.35   &   9.1 &   5.5 &   0   &   0   &   1.16    \\
BD +17$^{\circ}$3248    &   -2.08   &   10.2    &   3.8 &   0   &   0   &   2.75    \\
HD 6268     &   -2.42   &   3.2 &   2.4 &   0   &   0.01    &   2.36    \\
HD 128279   &   -2.48   &   0.6 &   2.9 &   0.02    &   0   &   1.99    \\
HD 126238   &   -1.98   &   2.1 &   3.3 &   0.01    &   0   &   0.34    \\
HD 122956   &   -1.95   &   7.2 &   5.1 &   0   &   0   &   0.67    \\
HD 6755     &   -1.68   &   6   &   4.2 &   0.01    &   0   &   0.40    \\
HD 115444   &   -2.90   &   6.0     &   4.1 &   0   &   0   &   1.46    \\
CS 31082-001    &   -2.9    &   50.9    &   3.8 &   0   &   0   &   0.68    \\
HD 122563   &   -2.72   &   0.3     &   4.1 &   0   &   0   &   1.89    \\
HD 175305   &   -1.48   &   2.1 &   2.2 &   0.02    &   0.01    &   1.31    \\
HD 94028    &   -1.75   &   2   &   4.9 &   2   &   0.2 &   0.57    \\
HD 76932    &   -1.18   &   5.1 &   3.9 &   0.4 &   0.9 &   0.63    \\
HD 107113   &   -0.79   &   8.7 &   2.4 &   0.4 &   0.5 &   0.37    \\
HD 2454     &   -0.72   &   1.3 &   2.5 &   10.9    &   0.6 &   0.58    \\
HD 16220    &   -0.56   &   1.6 &   2.4 &   2.3 &   1.3 &   0.11    \\

\hline

\end{tabular}
\end{minipage}
\end{table*}

\begin{figure*}
 \centering
 \includegraphics[width=1.0\textwidth,height=0.3\textheight]{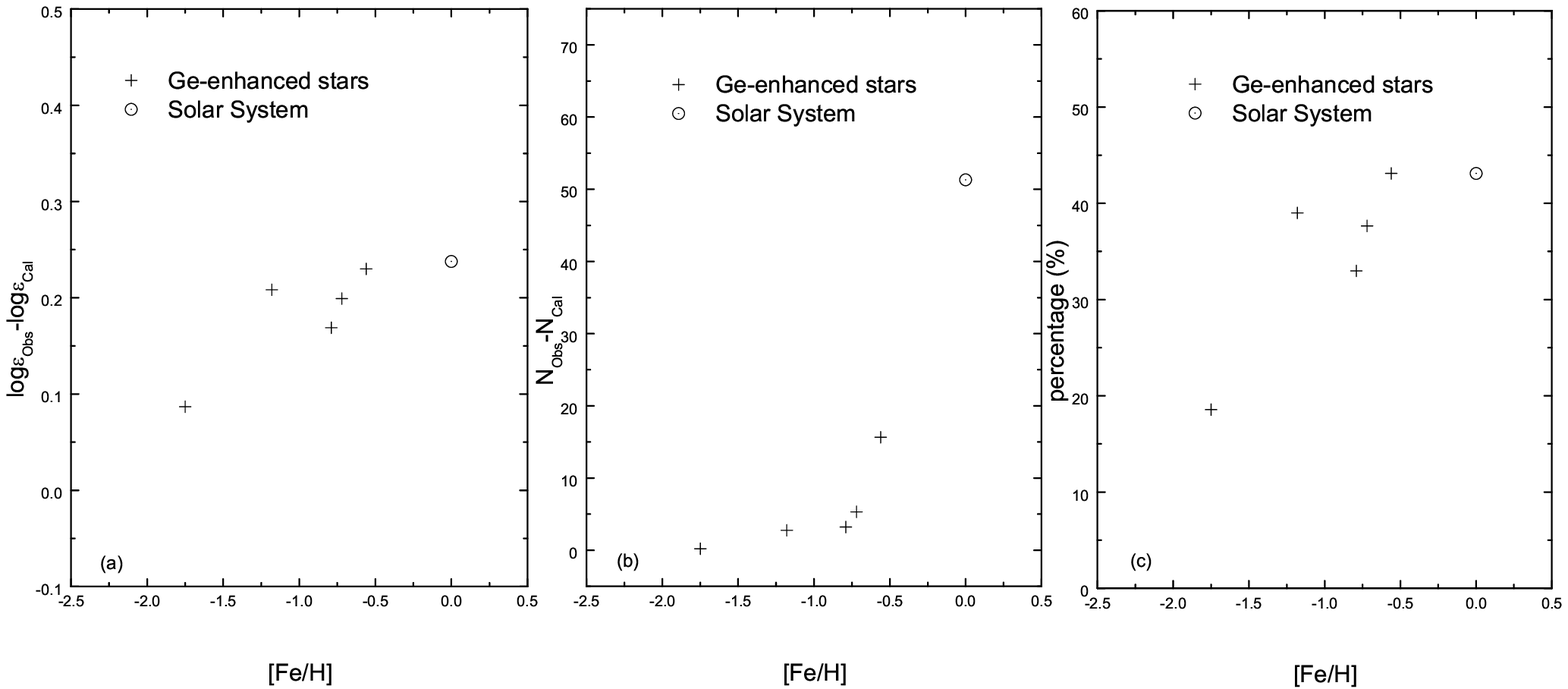}
\caption{The values of
$(log\varepsilon_{obs}-log\varepsilon_{cal})$ for the Ge-enhanced stars as the function of
metallicity [Fe/H] (a). The ``residual abundance" of Ge (=$N_{obs}-N_{cal}$)
as the function of metallicity [Fe/H] (b). The contributed percentage of residuals (=$1-N_{cal}/N_{obs}$) (c).}

\end{figure*}

\section{The contributions of the neutron-capture processes to Ge abundances}

Several previous studies on the abundances of the neutron-capture
elements have indicated that the observed abundances of the heavy
elements cannot be matched by only one neutron-capture process
\citep{Tr99,Al06,Li13b}. Based on the analysis of the observed
ratios [Ge/Fe] in the metal-poor stars, \citet{Pi10} estimated that
the contribution from the primary-like yields of the massive stars
to the solar Ge abundance is about 5\%$\sim$8\%. \citet{Tr04}
have found that the contributions from the weak s-process to the
abundances of metal-poor stars could be negligible, since the
secondary-like nature of major neutron source $^{ 22}Ne(\alpha,
n)^{25}Mg$.

Recently, \citet{Fr12} studied the impact of the initial rotation
rate on the production of several elements in the massive stars and
reported that rotating models can produce significant amounts of
elements up to Ba, even though in low-metallicity environments.
However, they found that the yields of light neutron-capture
elements have the secondary-like nature, which is different with the
primary nature for the r-process.

For exploring the astrophysical origin of elements Ge, we will
compare the observed abundances with the abundances contributed from
neutron-capture processes. In this case, the abundance of
neutron-capture element can be calculated as:


\begin{equation}
N_{i}(Z)= (C_{r,m}N_{i,r,m} + C_{r,w}N_{i,r,w} + C_{s,m}N_{i,s,m}+C_{s,w}N_{i,s,w})\times 10^{[Fe/H]}
\end{equation}

where $N_{i,r,m}$, $N_{i,r,w}$, $N_{i,s,m}$ and $N_{i,s,w}$ are the
abundances of the ith element produced by the main r-process, weak
r-process, main s-process and weak s-process respectively, which
have been normalized to corresponding abundances of the solar
system. The component coefficients $C_{r,m}$, $C_{r,w}$, $C_{s,m}$
and $C_{s,w}$ represent the relative contributions from the main
r-process, the weak-r process, the main s-process, the weak s
-process, respectively. Equation (5) has been used by \citet{Li13b}
to study the stellar abundances of the dwarf spheroidal galaxy. The
adopted metallicity-dependent abundance $N_{i,s,m}$ in Equation (5)
is taken from the main s-process abundance given by \citet{Bu01}
(see their Figure 1) and \citet{Bi10} (see their table B6). The
abundances of the weak s-process are taken from \citet{Tr04} for the
metal-rich stars. Obviously, there are some differences in the
abundance patterns of the weak s-process between the metal-poor
stars and the metal-rich stars. For the metal-poor stars, the
abundances of the weak s-process are taken from \citet{Fr12} (model
B4, $\upsilon_{ini}/\upsilon_{crit}=0.5$), since the calculated
results show that the abundance ratio [Ge/Zr]$\simeq-1.1$, which is
close to the average observed ratio of [Ge/Zr]$\simeq-1$.

Using Equation (5) and the observed data of the sample stars
\citep{Ho04,Ho07,Mi01,Ro10,Iv06,Co02,Co05,Ro12a,Ro12d,Si13,Hi02,We00},
we can obtain the best-fit $C_{r,m}$, $C_{r,w}$, $C_{s,m}$ and
$C_{s,w}$ by looking for the minimum $\chi^{2}$. In this step, the
observed Ge abundances have not been included, because Ge should not
be a "pure" neutron-capture element. There should have another
astrophysical origin to Ge, except for neutron-capture process,
for higher metallicity stars and the solar system. The component coefficients
and $\chi^{2}$ are listed in Table 4. From Table 4 we find that, for
the Ge-deficient stars(i.e., the most metal-poor stars from the sample
stars - see also Table 3), the weak s-process component coefficients
are smaller than 0.01 and the main s-process component coefficients
are smaller than 0.02, which are smaller than those of the r-processes
about two orders of magnitude. The calculated results imply that the
contributions of the s-process to the abundances of the Ge-deficient
stars are much smaller than those of the r-process, which is
consistent with the suggestion that the neutron-capture elements of
the Ge-deficient stars mainly come from the r-process. These results
can be naturally explained by the shorter lifetimes of massive
stars: massive stars in the early times of the Galaxy evolve
quickly, ending as SNe II producing r-process elements. On the other
hand, for the Ge-enhanced stars (the last 5 sample stars in Table
4), the two s-component coefficients are larger than those of the
Ge-deficient stars, which means that the contributions of the
s-process to the abundances of these stars become important. Based
on the calculations of galactic chemical evolution, \citet{Tr99}
have found that the main s-process contributions are significant
starting from [Fe/H]$\simeq-1.6$, since the longer lifetimes of low-
and intermediate-mass AGB stars. On the other hand, for the weak
s-process, little contribution is expected in halo stars, because of
the strong decrease in its efficiency with decreasing metallicity
\citet{Tr04}. Our calculated results are consistent with the
conclusions of \citet{Tr04}.

Based on the the discussions above, the Ge abundances in Ge-enhanced
stars would reflect the sum of contributions of the r-process,
s-process and the additional astrophysical process. Although the
observed Ge abundances have not been included in the fitted process,
it was possible to calculate the sum of Ge abundances of the four
components (i.e., neutron-capture process) using the component
coefficients and equation (5). To investigate whether the
contributions of the neutron-capture process can not interpret the
observed Ge abundances of the Ge-enhanced stars, we subtract the sum
of Ge abundances of the neutron-capture processes from the observed
Ge abundances. In this case, we define $\Delta\log\varepsilon(Ge)$
as: $\Delta\log\varepsilon(Ge)$
=$\log\varepsilon_{obs}-\log\varepsilon_{cal}$. The values of
$\Delta\log\varepsilon(Ge)$ for the Ge-enhanced stars as the
function of metallicity [Fe/H] are shown in Figure 7(a). From this
figure we can see that the values of $\Delta\log\varepsilon(Ge)$
increase with increasing of metallicity. There is a trend that the
values of $\Delta\log\varepsilon(Ge)$ reached the corresponding
value in the solar system at [Fe/H]=0. The calculated results imply
that, similar to the solar system, the contributions of the
neutron-capture processes cannot interpret the observed Ge
abundances of the Ge-enhanced stars. Figure 7(b) shows the Ge
``residual abundances" (=$N_{obs}-N_{cal}$) as the function of
metallicity [Fe/H]. The increased trend is visible. Figure 7(c) shows
the contribution percentages of the ``residual process" to the Ge
abundances of the Ge-enhanced stars. We can find that there is a
trend that the contribution percentages increase with increasing
metallicity. Recall that there is a trend in Figure 2 that the
ratios [Ge/H] increase from the line of the ratio of the weak
r-process star to the higher ratio started with [Fe/H]$\simeq -1.6$.
Based on the analysis above, we can find that the increased trend
should be attributed to the contributions of the weak s-process,
main s-process and the additional secondary process (i.e., the
yields increase with increasing initial metallicity). This secondary
process, which contributed Ge to ISM only for higher metallicity
([Fe/H]$\geq$-1.6), is responsible for the Ge ``residual abundance" in
the solar system and some higher metallicity stars. Note that the
contribution of the weak r-process to Ge abundances in metal-poor
stars could not be replaced by those of the ``residual process",
since the yields of the ``residual process" possess the secondary
nature.

\begin{figure*}
 \centering
 \includegraphics[width=1.0\textwidth,height=0.3\textheight]{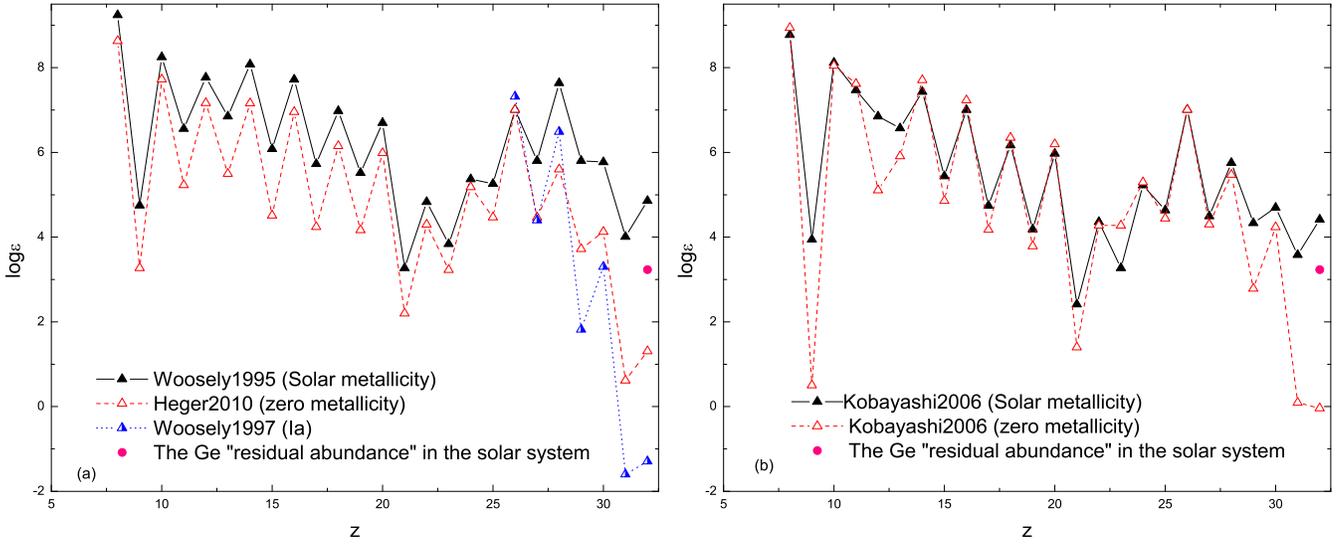}
\caption{The comparisons between averaged yields of massive stars
and the Ge `` residual abundance". The dashed lines show averaged
yields of the massive stars with zero metallicity presented by
\citet{He10}(a) and by \citet{Ko06}(b), which have been weighted by
the Salpeter IMF and scaled to the Fe abundance contributed from
massive stars in the solar system. The dotted line represent SNe Ia
yields presented by \citet{Wo97}, which have been normalized to the
Fe abundances contributed from SNe Ia in the solar system. The solid
lines show the averaged yields of the massive stars with the solar
metallicity ($Z= Z_{\odot}$) presented by \citet{Wo95}(a) and by
\citet{Ko06}(b), which also have been averaged by the Salpeter IMF
and scaled to the Fe abundance contributed from massive stars in the
solar system. Filled circle indicates the Ge ``residual abundance"
estimated in this work in the solar system.}


\end{figure*}

\section{INVESTIGATION OF ASTROPHYSICAL ORIGIN OF GE ``RESIDUAL ABUNDANCE"}

From Table 1 we can see that the contributed percentage of
Ge``residual abundance" to the solar system is about 41\%, which is
larger than the corresponding ``individual" values of main s-, weak
r-, main r-process. Considering the fitted results for metal-poor
stars, we find that Ge ``residual abundance" should appear with
higher metallicity [Fe/H]$\geq-1.6$. This means that the Ge yields
related to ``residual abundance" should have secondary nature. Since
Ge ``residual abundance" do not produced by the neutron-capture
processes, it must be produced by another process.

Historically, Ge (Z $= 32$) has been considered as a neutron-capture
element. It is noteworthy that the r-process abundances in the solar
system are derived through subtracting the s-process abundances from
total abundances of solar system \citep{Ar99}. If a heavy element
(e.g. Ge) was produced by the neutron-capture process partly, the
r-process abundance in solar system should have contained the
contribution of the other process (e.g. charged-particle synthesis).
We will investigate the astrophysical origin of Ge``residual
abundances" appeared at higher metalicity and in the solar system.

The element germanium is in the transition between charged-particle
synthesis of the iron-group elements and neutron-capture synthesis
of heavier elements. The substantial fractions of the iron-group
elements in the Solar system ($\sim33\%$ for Fe) were produced in
the massive stars under equilibrium or quasi-equilibrium conditions
\citep{Mi02}. However, the significantly heavier elements, such as
Sr and Zr, must have been produced by neutron-capture process.
Elements between these two extreme cases, such as Ga and Ge, may be
produced by each of these mechanisms. This implies that Ge
``residual abundance" in the solar system would be produced in the
massive stars. To test this point, the dashed lines in Figure 8(a) and
(b) show average yields of massive star with the zero
metallicity presented by \citet{He10} and \citet{Ko06} respectively,
which have been weighted by the IMF and scaled to the Fe abundances
contributed from massive stars in the solar system. The Ge
``residual abundance" in the solar system listed in Table 2 is also
shown in Figure 8(a) and (b) by filled circles. We can see that the
scaled Ge abundances are lower than the Ge ``residual abundance"
about $2.0\sim3.0$ dex, which means that the Ge ``residual
abundance" would not be produced as the primary-like yields in the
massive stars. The dotted line in Figure 8(a) shows the SNe Ia yields
presented by \citet{Wo97}, which have been normalized to the Fe
abundances contributed from SNe Ia in the solar system. The
normalized Ge abundance is lower than the Ge ``residual abundance"
in the solar system about 4 dex, which means that the Ge ``residual
abundance" in the solar system would not be produced in SNe Ia. The
solid lines in Figure 8(a) and (b) show averaged yields of the
massive stars with the solar metallicity ($Z=Z_{\odot}$) presented
by \citet {Wo95} and \citet{Ko06} respectively, which also have been
weighted by the IMF and scaled to the Fe abundances contributed from
massive stars in the solar system. We can see that, the scaled Ge
abundance is higher than Ge ``residual abundance" in the solar
system about $1.2\sim1.7$ dex. This should imply that, similar to
iron-group elements, the Ge ``residual abundance" in the solar
system is produced in the massive stars. The Ge yields related to
``residual abundance" should have secondary nature, which can be
observable for stars with higher metallicity ([Fe/H]$\geq-1.6$).

\section{CONCLUSIONS}

In our Galaxy, nearly all chemical evolution and nucleosynthetic
information is imprinted in the elemental abundances of stars with
various metallicities. In this respect, main r-process stars and
weak r-process stars are very significant, because their abundances
are polluted by only a few processes. On the other hand, the chemical
abundances of the metal-poor stars and metal-rich stars can also
provide important information. Our results can be summarized as
follows:

1. The Ge abundance in the metal-poor stars could not be produced as
the primary-like yields in the massive stars and should be produced
by the r-process. The Ge abundances of weak r-process and main
r-process are derived from the observed abundances of weak r-process
star and the main r-process star. We find that the observed relation
of [Ge/H] and [Eu/H] are close to the r-process abundance ratio in
the solar system and the Ge abundances of metal-poor stars can be
fitted by combined contributions of the weak r- and main
r-processes.

2. Because both iron group elements and the weak r-process elements
would be produced in the massive stars at low metallicity, the
observed correlation of the [Ge/H] with the [Fe/H] could be
understood. Furthermore, since the Ge abundances dominantly come
from the weak r-process for the most Ge-deficient stars, the
increasing of [Eu/Fe] directly leads to the observed decline of
[Ge/Eu].

3. For the Ge-enhanced stars, there is a increased trend in [Ge/H]
started from the line of abundance ratio of the weak r-process star
at [Fe/H]$\simeq$-1.6 to the line of [Ge/H]=[Fe/H], which should
be attributed to the contributions of the weak s-process, main
s-process and the secondary process occurred in the massive stars.

4. The contributed percentages of main r-, weak r-, main s-and weak
s-processes for Ge to the abundance of the solar system are about
0.32\%, 3.25\%, 12.02\% and 43.03\% respectively, which means that
the contributed percentage of ``residual abundance" to the solar
system is about 41\%. Our investigations imply that the Ge
``residual abundance" would be produced as the secondary-like yields
in the massive stars.

Our calculated results should give the constraints on models of the
r-process that produced Ge element in the Galaxy. We hope that the
results here will present interesting information to more complete
models of Galactic chemical evolution. A more precise knowledge
about the Ge abundances in metal-poor stars and in population I
stars is needed.

\section*{Acknowledgments}

We thank the referee for an extensive and helpful review, containing
very relevant scientific suggestions that greatly improved this
paper . This work has been supported by the National Natural Science
Foundation of China under 11273011, U1231119, 10973006 and 11003002,
the Natural Science Foundation of Hebei Provincial Education
Department under grant Z2010168, XJPT002 of Shijiazhuang University,
the Natural Science Foundation of Hebei Province under Grant
A2011205102, A2011210017, and the Program for Excellent Innovative
Talents in University of Hebei Province under Grant CPRC034.

\bsp

\label{lastpage}

\end{document}